\documentclass[journal]{IEEEtran}
\usepackage{amsthm,amssymb,graphicx,multirow,amsmath,color,amsfonts}%,ulem}
\usepackage[update,prepend]{epstopdf}
\usepackage[noadjust]{cite}
\usepackage[latin1]{inputenc}
\usepackage{tikz}
\usetikzlibrary{arrows,calc}
\usepackage{bbm}
\usepackage{pdfpages}
\usepackage{tabulary}
\usepackage{multirow}
\usepackage{comment}

\def\nbo{{\mathbf{o}}}

\def\nbx{{\mathbf{x}}}
\def\nby{{\mathbf{y}}}

\def\nb0{{\mathbf{0}}}
\def\nb1{{\mathbf{1}}}

\def\nbX{{\mathbf{X}}}
\def\nbY{{\mathbf{Y}}}

\def\ncalA{{\mathcal{A}}}
\def\ncalB{{\mathcal{B}}}

\def\ncalL{{\mathcal{L}}}

\def\nbbE{{\mathbb{E}}}

\def\nbbP{{\mathbb{P}}}

\newtheorem{lemma}{Lemma}

\newtheorem{theorem}{Theorem}

\newtheorem{remark}{Remark}

\allowdisplaybreaks
\begin{document}
\title{3GPP-inspired Stochastic Geometry-based Mobility Model for a Drone Cellular Network}
\author{Morteza Banagar and Harpreet S. Dhillon
\thanks{The authors are with Wireless@VT, Department of ECE, Virginia Tech, Blacksburg, VA (email: \{mbanagar, hdhillon\}@vt.edu). The support of the US NSF (Grant CNS-1617896) is gratefully acknowledged.}}

\maketitle

\begin{abstract}
This paper deals with the stochastic geometry-based characterization of the time-varying performance of a drone cellular network in which the initial locations of drone base stations (DBSs) are modeled as a Poisson point process (PPP) and each DBS is assumed to move on a straight line in a random direction. This drone placement and trajectory model closely emulates the one used by the third generation partnership project (3GPP) for drone-related studies. Assuming the nearest neighbor association policy for a typical user equipment (UE) on the ground, we consider two models for the mobility of the serving DBS: (i) {\em UE independent model}, and (ii) {\em UE dependent model}. Using displacement theorem from stochastic geometry, we characterize the time-varying interference field as seen by the typical UE, using which we derive the time-varying coverage probability and data rate at the typical UE. We also compare our model with more sophisticated mobility models where the DBSs may move in nonlinear trajectories and demonstrate that the coverage probability and rate estimated by our model act as lower bounds to these more general models. To the best of our knowledge, this is the first work to perform a rigorous analysis of the 3GPP-inspired drone mobility model and establish connection between this model and the more general non-linear mobility models.
\end{abstract}


\section{Introduction} \label{sec:Intro}
With the increasing maturity of {\em drone technology}, wireless networks are all set to undergo a major transformation from predominantly terrestrial networks to the ones that will have an elaborate and dynamic aerial component in the form of drone networks \cite{J_Chandrasekharan_Designing_2016, J_Mozaffari_Tutorial_2018}. The flexibility of on-demand deployments of DBSs has added an entirely new dimension to the system design that was not present in the traditional terrestrial networks. However, in order to appropriately use this additional dimension, one needs to carefully analyze the performance of the resulting network in which base stations (BSs) are mobile. Given the expected irregularity in the deployment and trajectories of drones, it is not unreasonable to expect that tools from stochastic geometry could be leveraged to obtain system design insights, which is the key focus of this paper. We also show that the approach taken by 3GPP for the mobility analysis of DBSs is surprisingly amenable to the stochastic geometry treatment, which will form the basis for our drone mobility model \cite{3gpp_36777}.

\emph{Related Works.}
The past decade has seen the emergence of stochastic geometry as a powerful tool for the performance analysis of communication networks that perfectly complements the traditional simulation-driven approaches \cite{B_Haenggi_Stochastic_2012, J_Dhillon_Modeling_2012}. There has been recent interest in applying these tools to the analysis of drone networks as well. While the initial studies, e.g., \cite{J_Chetlur_Downlink_2017, C_Chetlur_Downlink_2016}, were focused on static drone networks, mobility has also been considered in the more recent works, e.g., \cite{J_Enayati_mobile_2018, J_Hayajneh_Performance_2018, C_Fotouhi_Dynamic_2016, J_Sharma_Coverage_2019, J_Sharma_Random_2019}. We discuss these works in more detail next.

Downlink coverage probability of a finite static network of DBSs is characterized in \cite{J_Chetlur_Downlink_2017, C_Chetlur_Downlink_2016}. In these works, the authors modeled a finite network of DBSs as a uniform binomial point process (BPP) and derived the exact expression for the coverage probability of the network. Using the results of \cite{J_Chetlur_Downlink_2017}, the authors in \cite{J_Enayati_mobile_2018} considered a mobile network of drones and examined two trajectory processes for the movement of drones in order to preserve the BPP distribution of drones at any time instance, which leads to an almost uniform coverage. Analysis of a drone-based small cell network is done in \cite{J_Hayajneh_Performance_2018}, in which the authors considered a post-disaster situation where drones are deployed based on cluster point process in the vicinity of a damaged BS. The performance benefits of dynamically repositioning DBSs are explored in \cite{C_Fotouhi_Dynamic_2016}. The authors developed algorithms to autonomously control the movement of the drones as users move on the ground. The authors in \cite{J_Sharma_Coverage_2019} considered a finite 3D network of mobile drones and proposed a mixed mobility model for the movement of interfering drones, while the serving drone hovers at a fixed height above the location of the UE. In their model, interfering drones move vertically and horizontally based on the random waypoint (RWP) and random walk (RW) mobility models, respectively. In their other work \cite{J_Sharma_Random_2019}, the authors assumed that the serving drone can also move based on the same mobility model as the interfering drones and analyzed the coverage probability of the network. Based on these prior arts and motivated by the simulation models of 3GPP (to be described in the next section), we analyze the coverage probability and the received rate of the UE for a mobile network of DBSs. We will elaborate more on our contributions next.

\emph{Contributions and Outcomes.}
Inspired by the simulation model considered in the drone-related studies by 3GPP \cite{3gpp_36777}, we model the initial locations of a mobile network of DBSs operating at a constant height by a homogeneous PPP. Considering the nearest DBS to the typical ground UE as the serving DBS, we propose two service models for the mobility of the serving DBS, i.e., (i) {\em UE independent model}, where the serving DBS moves in a random direction, and (ii) {\em UE dependent model}, where the serving DBS moves towards the UE at a constant height and hovers when it reaches above the location of the UE. All the other DBSs are regarded as interfering DBSs and we assume that they move on straight lines and in random directions, independent of the movement of the serving DBS. With the constant velocity assumption and conditioned on the nearest DBS distance, we characterize the time-varying interference field as seen by the typical UE and demonstrate for the second service model that the locations of all other DBSs form an inhomogeneous PPP as they move randomly. Using this, we then characterize the time-varying coverage and rate for the typical UE for both the service models. We also compare our proposed straight-line mobility model with the one where DBSs can also change their directions and demonstrate that the proposed model provides lower bounds on the time-varying coverage and rate for these more complicated non-linear models. To the best of our understanding, this is the first paper that provides a comprehensive analysis for the 3GPP-inspired drone mobility model and connects it to more complicated non-linear mobility models.

%%%%%%%%%%%%%%%%%%%%%%
\section{System Model} \label{sec:SysMod}
\begin{figure}[t!]
    \centering
    \includegraphics[width=0.9\columnwidth]{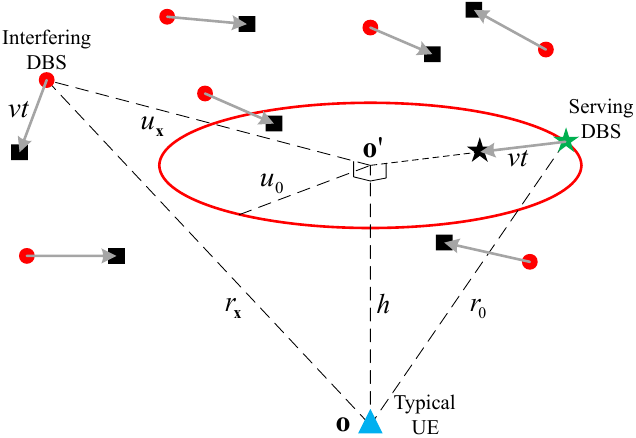}
    \caption{An illustration of the system model and the movement scenario. Green and black stars, red circles, black squares, and blue triangle represent serving DBS, displaced serving DBS, interfering DBSs, displaced interfering DBSs, and the typical UE, respectively.} \vspace{-0.3cm}
    \label{Fig:SysMod}
\end{figure}

%%%%%%%%%%%%%%%%%%%%%%
We consider a network of mobile DBSs which are deployed to serve UEs on the ground. The DBSs are located at height $h$ and the projections of the DBS locations on the ground are initially distributed based on a homogeneous PPP $\Phi_{\rm D}(0)$ with density $\lambda_0$. UEs are distributed on the ground based on another homogeneous PPP $\Phi_{\rm U}$. Note that the origin ${\bf o} = (0,0,0)$ of the 3D coordinate system is located on the ground and we refer to the $z=h$ plane as the DBS plane throughout this paper. We also denote the projection of the origin onto the DBS plane by ${\bf o'} = (0, 0, h)$. The analysis will be performed for the {\em typical} UE placed at ${\bf o}$. As shown in Fig. \ref{Fig:SysMod}, the distances of a DBS at time $t$ located at $\nbx(t)\in\Phi_{\rm D}(t)$ from $\bf{o'}$ and $\bf{o}$ are denoted by $u_\nbx(t) = \|\nbx(t) - \bf{o'}\|$ and $r_\nbx(t) = \sqrt{u_\nbx(t)^2+h^2}$, respectively. Moreover, the location of the nearest DBS to $\bf{o'}$ and its corresponding distance at time $t$ are denoted by $\nbx_0(t)$ and $u_0(t)$, respectively. Thus, the distance of the closest DBS to $\bf{o}$ at time $t$ is $r_0(t) = \sqrt{u_0(t)^2+h^2}$. For notational simplicity of the initial distances, we drop the time argument $t$ when $t=0$, i.e., $u_0 \triangleq u_0(0)$, $r_0 \triangleq r_0(0)$, $u_\nbx \triangleq u_\nbx(0)$, and $r_\nbx \triangleq r_\nbx(0)$.

In the 3GPP simulation model for the placement and trajectories of drones and their mobility evaluations, it is assumed that drones start their movement at randomly selected locations in the network. They then move at a constant velocity and height in straight lines and in uniformly random directions for the entire duration of the simulation \cite{3gpp_36777}. In this paper, we assume that the typical UE connects to its nearest DBS. We denote this DBS as the serving DBS for the typical UE and all the other DBSs are regarded as interfering DBSs. Motivated by the 3GPP mobility model, we assume that interfering DBSs move in straight lines with a constant velocity of $v$ and in random directions independent of the other DBSs. For the serving DBS, however, we consider two movement scenarios as follows.
\begin{enumerate}
\item UE independent model: The serving DBS moves with a constant velocity of $v$ in a random direction, independently of the UE location.
\item UE dependent model: The serving DBS moves with a constant velocity of $v$ in the DBS plane towards $\bf{o'}$ and stops at this location.
\end{enumerate}
Note that the first service model above is similar to the movement scenario for the interfering DBSs. Due to the random direction of movement of the serving DBS in the first service model, the closest DBS to the typical UE will not remain the same over time, which causes a handover to occur. However, in the second service model, since the serving DBS moves in a way to minimize its distance from the typical UE, as long as all DBSs move with the same velocity $v$, this service model captures the best-case scenario for the typical UE in the sense of achieving the minimum distance between the serving DBS and the typical UE. Hence, handover will not occur in the second service model. Moreover, since we are not considering any dependency across the user locations, DBS trajectories will be independent of each other in both service models.

The received signal-to-interference-plus-noise ratio (${\rm SINR}$) at time $t$ is defined as
\begin{equation} \label{Eq:SINR}
{\rm SINR}(t) = \frac{P h_0(t) r_0(t)^{-\alpha}}{I(t) + N_0},
\end{equation}
where $P$ is the DBS transmit power, which is considered to be equal for all DBSs at all times, $h_0(t)$ represents the small-scale fading gain between the typical UE and the serving DBS, $\alpha>2$ is the path loss exponent, $N_0$ is the thermal noise power at the typical UE, and $I(t)$ is the interference power defined as $I(t)=\sum_{\nbx(t) \in \Phi_{\rm D}'(t)}P h_\nbx(t) r_\nbx(t)^{-\alpha}$, where $\Phi_{\rm D}'(t) \equiv \Phi_{\rm D}(t)\backslash \nbx_0(t)$ represents the PPP of interference field and $h_\nbx(t)$ denotes the small-scale fading gains between the typical UE and the interfering DBSs. We assume Rayleigh fading with a mean power of $1$ for DBS-UE channels, and thus, we have $h_0(t) \sim {\rm exp}(1)$ and $h_\nbx(t) \sim {\rm exp}(1)$. Note that the second service model implies that $u_0(t) = [u_0 - vt]^+$, where $[a]^+=a$ if $a\geq 0$ and $[a]^+=0$ otherwise.

In order to analyze the proposed network, we introduce two ${\rm SINR}$-based metrics, namely time-varying coverage probability and rate. $P_{\rm C}(\gamma, t)$ is defined as the probability that ${\rm SINR}(t)$ at the typical UE is greater than some predetermined constant threshold $\gamma$ and $R(t)$ is given as $R(t) = \nbbE[\log\left(1 + {\rm SINR}(t)\right)]$, where the expectation is taken over the PPP $\Phi_{\rm D}$ and the trajectories.

%%%%%%%%%%%%%%%%%%%%%%
\section{Coverage Probability and Rate} \label{sec:ASE}
\begin{figure}[t!]
    \centering
    \includegraphics[width=1\columnwidth]{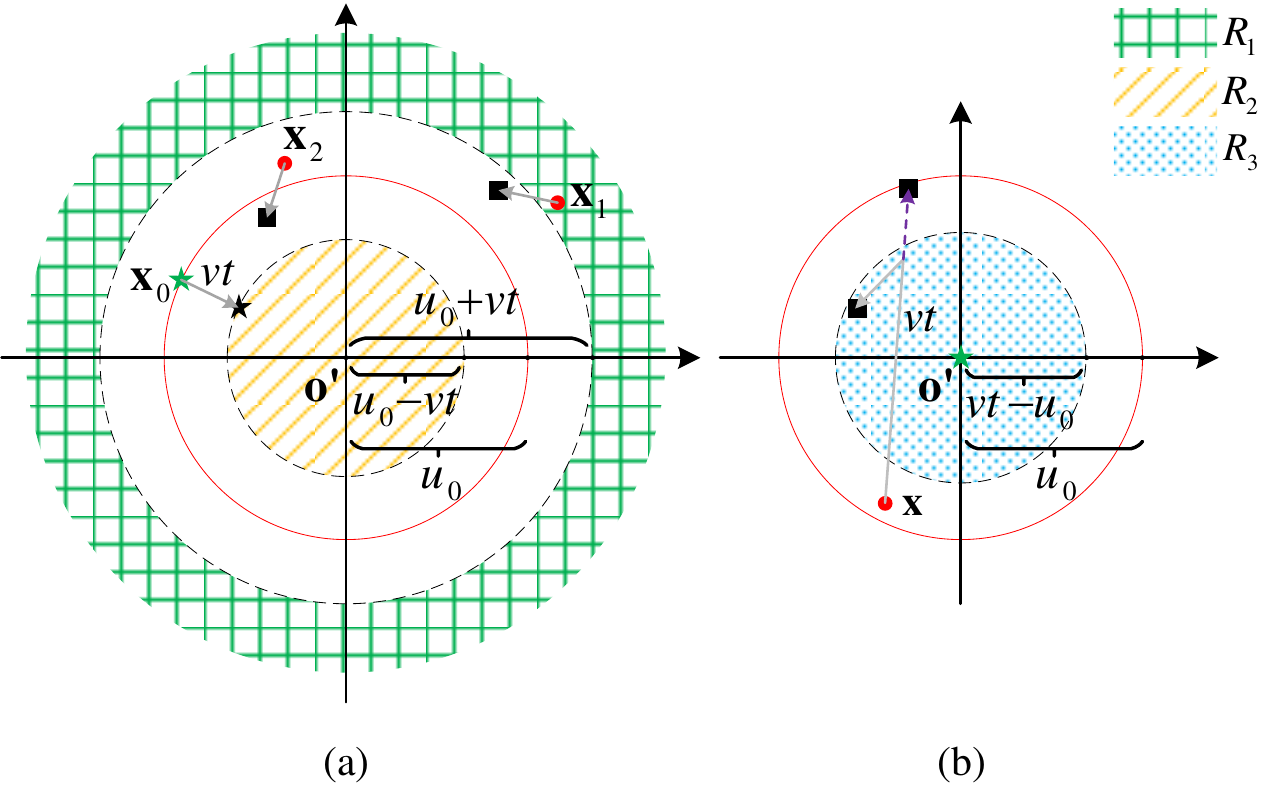}
    \caption{An illustrative explanation of the network density for different regions. Green and black stars, red circles, and black squares represent serving DBS, displaced serving DBS, interfering DBSs, and displaced interfering DBSs, respectively. (a) Serving DBS is moving from $\nbx_0$ towards $\nbo'$, and (b) serving DBS is hovering at $\nbo'$.} \vspace{-0.3cm}
    \label{Fig:Remark2}
\end{figure}

%%%%%%%%%%%%%%%%%%%%%%
Prior to the analysis of the coverage probability and rate, we need to first characterize the density of the network of interfering DBSs. We start our analysis by considering the first service model, i.e., the UE independent mobility model for the serving DBS. The following lemma forms the basis of our analysis.
\begin{lemma} \label{lem:NoExclusion}
Let $\Phi_{\rm D}$ be a PPP with density $\lambda_0$. If all the points of $\Phi_{\rm D}$ are displaced independently of each other and their displacements are identically distributed, then the displaced points form another PPP $\Phi'_{\rm D}$ with the same density $\lambda_0$.
\end{lemma}
\begin{IEEEproof}
Since the distance traveled by each DBS after time $t$ is a constant $vt$, which is also independent of the original locations of the DBSs in $\Phi_{\rm D}$, the lemma follows directly from the displacement theorem \cite{B_Haenggi_Stochastic_2012}.
\end{IEEEproof}
Lemma \ref{lem:NoExclusion} states that if all the DBSs (including the serving DBS) are displaced randomly based on our straight-line mobility model, then the spatial distribution of the drones will not change. Therefore, for the first service model, the network of interfering DBSs follows an inhomogeneous PPP with the density given by
\begin{equation} \label{LambdaServiceModel1}
\lambda(u_\nbx, u_0)=\left\{\begin{matrix}
\lambda_0 & u_\nbx > u_0\\ 
0 & u_\nbx \leq u_0
\end{matrix}.\right.
\end{equation}
Note that although the value of $u_0$ in \eqref{LambdaServiceModel1} varies over time, its distribution does not change. The interference field in the second service model will be an inhomogeneous PPP as well. To clarify this, let us assume that the serving DBS is initially located at distance $u_0$ from $\bf{o'}$. It is clear from our construction that initially there is no other DBS within a disc of radius $u_0$ centered at $\bf{o'}$. We denote this \emph {exclusion zone} by $b({\bf o'}, u_0)$. Hence, the density of the network of interfering DBSs is initially given as \eqref{LambdaServiceModel1}. As the serving DBS moves towards $\bf{o'}$, interfering DBSs can enter the initial exclusion zone, and thus, their density will change. In the next lemma, we will characterize the density of the network of interfering DBSs for the second service model.
\begin{lemma} \label{lem:MainDensity}
In the second service model, if the interfering DBSs move in independent uniformly random directions and in straight lines with the same velocity as the serving DBS, then the locations of interfering DBSs will be distributed as an inhomogeneous PPP with density
\begin{align} \label{MainLambda1}
&\lambda(u_\nbx, u_0, t) =\nonumber\\
&\lambda_0
  \begin{cases}
     1 & u_0 + vt \leq u_\nbx\\
     \frac{1}{\pi}\cos^{-1}\left(\frac{u_0^2 - u_\nbx^2 - v^2 t^2}{2u_\nbx v t}\right) & |u_0 - vt| \leq u_\nbx \leq u_0 + vt\\
     {\bf 1}\left(t>\frac{u_0}{v}\right) & 0 \leq u_\nbx \leq |u_0 - vt|
  \end{cases},
\end{align}
where ${\bf 1}(.)$ is the indicator function.
\end{lemma}
\begin{IEEEproof}
Since the initial density of the network of interfering DBSs as given in \eqref{LambdaServiceModel1} is inhomogeneous and the displacements are independent of each other, displacement theorem states that the resulting network will also be an inhomogeneous PPP \cite{B_Haenggi_Stochastic_2012}. In order to find the density of the network of displaced DBSs, we need to first characterize the displacement kernel $\rho(\nbX, .)$ of our mobility model, i.e., the distribution of the displaced location of a point at $\nbX$. Assume that an interfering DBS is initially located at $\nbX$, which after a displacement of $vt$ in a random direction of $\Theta$, lands on another location $\nbY$. Let the distance from the DBS to $\nbo'$ before and after the displacement be $u_\nbx$ and $u_\nby$, respectively. Using the cosine law, we have
\begin{align*}
u_\nby^2 = u_\nbx^2 + v^2t^2 - 2u_\nbx vt \cos(\Theta).
\end{align*}
Since $\Theta$ is uniformly distributed in $[0, 2\pi)$, we can write the distribution of the new locations of DBSs using the basic transformation of random variables as
\begin{equation*} \label{Eq2:Lemma1}
\rho(u_\nbx, u_\nby) = \frac{2u_\nby}{\pi\sqrt{\left(u_\nby^2-(u_\nbx - vt)^2\right)\left((u_\nbx + vt)^2-u_\nby^2\right)}},
\end{equation*}
when $|u_\nbx-vt|\leq u_\nby \leq u_\nbx+vt$ and zero otherwise. These conditions are the well-known triangle inequalities, which in terms of $u_\nbx$ and by taking the exclusion zone into account can be rewritten as $\max\{u_0, |u_\nby-vt|\}\leq u_\nbx \leq u_\nby+vt$. Now, displacement theorem gives the density $\lambda(u_\nby, u_0, t)$ of the displaced network in polar coordinates as
\begin{align*}
& 2\pi u_\nby\lambda(u_\nby, u_0, t) \\
&= 2\pi\int_{u_0}^{\infty}\lambda_0 \rho(u_\nbx, u_\nby) u_\nbx {\rm d}u_\nbx \\
&= \int_{\max\{u_0, |u_\nby-vt|\}}^{u_\nby+vt} \frac{4\lambda_0 u_\nbx u_\nby \,\, {\rm d}u_\nbx}{\sqrt{\left(u_\nby^2-(u_\nbx - vt)^2\right)\left((u_\nbx + vt)^2-u_\nby^2\right)}}\\
&=2\lambda_0 u_\nby\cos^{-1}\left( \frac{\left( \max\{u_0, |u_\nby-vt|\} \right)^2 - (u_\nby^2 + v^2t^2)}{2vt u_\nby} \right),
\end{align*}
which after some simplifications gives \eqref{MainLambda1} and the proof is complete.
\end{IEEEproof}
From Lemma \ref{lem:MainDensity}, we give the following remarks.
\begin{remark} \label{Remark1}
Note that \eqref{MainLambda1} is continuous at the boundaries, i.e., at $u_\nbx = |u_0 \pm vt|$. Moreover, as $u_0 \to 0$ or $t \to \infty$, the underlying inhomogeneous PPP of the interferers will become homogeneous (meaning $\lambda(u_\nbx, u_0 \to 0, t) = \lambda(u_\nbx, u_0, t \to \infty) = \lambda_0$). Finally, equations \eqref{MainLambda1} and \eqref{LambdaServiceModel1} will become identical for $t = 0$ or $u_0 = 0$, as expected.
\end{remark}
\begin{remark} \label{Remark1_1}
Another way of characterizing the interference field of the second service model can be summarized as follows. According to Lemma \ref{lem:NoExclusion}, if there was no exclusion zone, the density of the network of DBSs (including the serving DBS) would be $\lambda_0$ as they move independently based on our straight-line mobility model. With the introduction of the exclusion zone, the density of the network can be viewed as the superposition of two parts: (i) density imposed by the exclusion zone, and (ii) density of the interference field. Hence, computing the network density imposed by the exclusion zone gives us the density of the network of interfering DBSs as well.
\end{remark}
\begin{remark} \label{Remark2}
Intuitively, no DBS in the region $R_1=\{u_\nbx \geq u_0 + vt\}$ (shaded as green in Fig. \ref{Fig:Remark2} (a)) can enter the original exclusion zone $b({\bf o'}, u_0)$ until time $t$, and thus, $\lambda(u_\nbx, u_0, t) = \lambda_0$. For the region $R_2=\{0 \leq u_\nbx \leq u_0 - vt | vt \leq u_0\}$ (shaded as orange in Fig. \ref{Fig:Remark2} (a)), the serving DBS is still in motion towards $\nbo'$ and since no interfering DBS exists in $R_2$, we have $\lambda(u_\nbx, u_0, t) = 0$. Finally for $R_3=\{0 \leq u_\nbx \leq vt - u_0 | vt \geq u_0\}$ (shaded as blue in Fig. \ref{Fig:Remark2} (b)), we can use the interpretation of the network density mentioned in Remark \ref{Remark1_1} to first calculate the density imposed by the exclusion zone and then subtract it from $\lambda_0$ to get the density of the interfering DBSs. Observe that DBSs initially inside the exclusion zone will not remain in the $R_3$ after the displacement of $vt$. Hence, the density imposed by the exclusion zone in $R_3$ is zero, which gives $\lambda(u_\nbx, u_0, t) = \lambda_0$.
\end{remark}

\begin{remark} \label{Remark3}
The analysis of other mobility models, such as RW and RWP mobility models \cite{J_Sharma_Random_2019}, in which drones can change their directions during their flights, is more complex and requires careful characterization of the resulting point processes. However, it turns out that it is possible to make a fairly general statement about the performance comparison of these more general models with our straight-line mobility model, which we do now. Based on Remark \ref{Remark1_1} for $t \geq \frac{u_0}{v}$, if a DBS at $\nbx$ follows our straight-line mobility model, it will not fall into the region $R_1$ when it travels a distance of $vt$ (see the purple dotted arrow in Fig. \ref{Fig:Remark2} (b)). On the other hand, using RW or RWP mobility models which allow direction changes on flights, a DBS that is initially at $\nbx$ could possibly fall into $R_1$ after a displacement of $vt$. This means that the impact of exclusion zone is not zero in these mobility models, and thus, the density of the network of interfering DBSs will be less than $\lambda_0$ in $R_1$. Hence, compared to our straight-line mobility model, less interferers will be present in the vicinity of the typical UE when DBSs move based on RW or RWP mobility models. As a result, the received rate at the typical UE will be higher when DBSs can change their directions while moving compared to our straight-line mobility model. Note that we were also able to establish these performance bounds analytically but could not include them here due to lack of space. They will be included in the expanded journal version of the paper.
\end{remark}

We are now equipped to analyze the distribution of ${\rm SINR}(t)$ and the rate of the typical UE. Theorem \ref{theo:ASE} states the main result of this paper.
\begin{theorem} \label{theo:ASE}
In the second service model, the time-varying coverage probability and the instantaneous received rate at the typical UE are given as
\begin{align}
P_{\rm C}(\gamma, t) &= \int_0^{vt} 2\pi\lambda_0 u_0 {\rm e}^{-\pi \lambda_0 u_0^2 -\frac{\gamma h^{\alpha}}{P}N_0 - 2\pi\lambda_0 \ncalA}\,{\rm d}u_0 \,+ \nonumber\\
&\hspace{-1cm}\int_{vt}^\infty 2\pi\lambda_0 u_0 {\rm e}^{-\pi \lambda_0 u_0^2 -\frac{\gamma \left((u_0-vt)^2+h^2\right)^{\alpha/2}}{P}N_0 - 2\pi\lambda_0 \ncalB}\,{\rm d}u_0, \label{MainCoverage}\\
R(t) &= \int_0^\infty \frac{P_{\rm C}(\gamma, t)}{1+\gamma}\,{\rm d}\gamma, \label{MainRates}
\end{align}
where
\begin{align*}
\ncalA &= \int_0^\infty g_1(u_\nbx) \,{\rm d}u_\nbx \,- \\
&\int_{vt-u_0}^{vt+u_0} g_1(u_\nbx)\frac{1}{\pi}\cos^{-1}\left(\frac{-u_0^2 + u_\nbx^2 + v^2 t^2}{2u_\nbx v t}\right) \,{\rm d}u_\nbx, \\
\ncalB &= \int_{u_0+vt}^\infty g_2(u_\nbx) \,{\rm d}u_\nbx \,+ \\
&\int_{u_0-vt}^{u_0+vt} g_2(u_\nbx)\frac{1}{\pi}\cos^{-1}\left(\frac{u_0^2 - u_\nbx^2 - v^2 t^2}{2u_\nbx v t}\right) \,{\rm d}u_\nbx, \\
g_1(u_\nbx) &= \frac{u_\nbx}{1 + \frac{1}{\gamma}\left(\frac{u_\nbx^2+h^2}{h^2}\right)^{\alpha/2}}, \nonumber\\
g_2(u_\nbx) &= \frac{u_\nbx}{1 + \frac{1}{\gamma}\left(\frac{u_\nbx^2+h^2}{(u_0-vt)^2+h^2}\right)^{\alpha/2}}.
\end{align*}
\end{theorem}
\begin{IEEEproof}
Conditioning on the location of the serving DBS, we write the coverage probability for the ${\rm SINR}$ threshold $\gamma$ as
\begin{align*}
P_{\rm C}(\gamma, t\bigr\rvert\nbx_0(t)) &= \nbbP\left[{\rm SINR}(t) \geq \gamma \bigr\rvert\nbx_0(t) \right]\\
&\hspace{-1cm}= \mathbb{E}\left[\nbbP\left[ h_0(t) \geq \frac{\gamma r_0^\alpha(t)(N_0+I(t))}{P}\middle|\nbx_0(t), I(t) \right]\right]\\
&\hspace{-1cm}= \exp\left[-\frac{\gamma r_0^\alpha(t)}{P}N_0\right] \ncalL_{I(t)}\left(s\bigr\rvert\nbx_0(t)\right)\biggr|_{s=\frac{\gamma r_0^\alpha(t)}{P}},
\end{align*}
where we have used the Rayleigh fading assumption in the last equation. Note that the expectation is taken over $I(t)$ and $\ncalL_{I(t)}(s\bigr\rvert\nbx_0(t)) = \nbbE\left[{\rm e}^{-sI(t)}\bigr\rvert\nbx_0(t)\right]$ is the conditional Laplace transform of interference at time $t$. We now determine the conditional Laplace transform of $I(t)$ as follows.
\begin{align*}
\ncalL_{I(t)}(s\bigr\rvert\nbx_0(t)) &= \nbbE\left[\exp\left[-s\!\!\!\!\!\!\!\!\sum_{\nbx(t) \in \Phi_{\rm D}'(t)}\!\!\!\!\!\!\!\!\!P h_\nbx(t) r_\nbx(t)^{-\alpha}\right] \middle|u_0(t) \right] \\
&\hspace{-1cm}\overset{(a)}{=} \nbbE\left[ \prod_{\nbx(t) \in \Phi_{\rm D}'(t)} \frac{1}{1 + sP(u_\nbx^2(t)+h^2)^{-\alpha/2}} \middle|u_0(t) \right] \\ 
&\hspace{-1cm}\overset{(b)}{=} \exp\left[-2\pi\int_{0}^\infty\frac{u_\nbx(t) \lambda(u_\nbx, u_0, t)}{1 + \frac{1}{sP}(u_\nbx^2(t)+h^2)^{\alpha/2}}\,{\rm d}u_\nbx(t)\right],
\end{align*}
where (a) follows from the moment generating function (MGF) of the exponential distribution and (b) results from the probability generating functional (PGFL) of a PPP. Hence, deconditioning on the serving DBS distance $u_0(t)$, the time-varying coverage probability can be written as
\begin{align*}
P_{\rm C}(\gamma, t) =\, &\int_0^\infty 2\pi\lambda_0 u_0 {\rm e}^{-\pi \lambda_0 u_0^2} \times {\rm e}^{-\frac{\gamma \left(u_0^2(t)+h^2\right)^{\alpha/2}}{P}N_0} \times \nonumber\\
&\exp\Big(- 2\pi{\displaystyle \int_{0}^\infty}\frac{u_\nbx \lambda(u_\nbx, u_0, t)}{1 + \frac{1}{\gamma}\left(\frac{u_\nbx^2+h^2}{u_0^2(t)+h^2}\right)^{\alpha/2}}\,{\rm d}u_\nbx\Big) \,{\rm d}u_0,
\end{align*}
where we have removed the argument of $u_\nbx(t)$ for simplicity. Applying $u_0(t) = [u_0 - vt]^+$, we end up with \eqref{MainCoverage}.
\begin{figure}[t!]
    \centering
    \includegraphics[width=0.85\columnwidth]{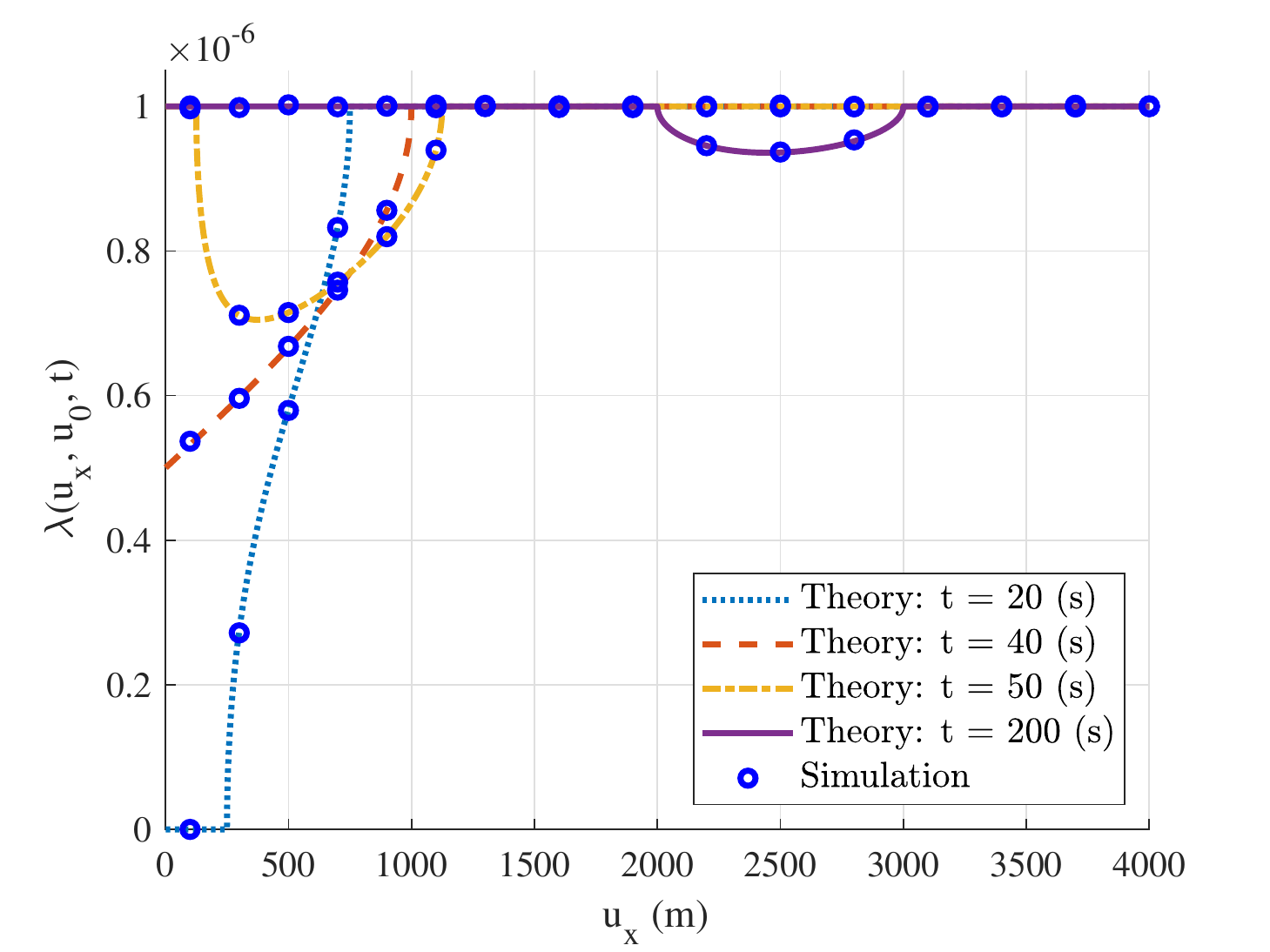}
    \caption{Density of the network of interfering DBSs when the serving DBS moves based the second service model. The exclusion zone radius is $u_0 = 500~{\rm m}$. The homogenization of the interference field with time is clear.}\vspace{-0.3cm}
    \label{Fig:DensityPlot}
\end{figure}

We now determine the time-varying rate for the typical UE. By definition, we can write the time-varying rate as
\begin{align*}
R(t) &= \nbbE[\log\left(1 + {\rm SINR}(t)\right)] = \int_0^\infty \log(1+\gamma)f_\Gamma(\gamma; t)\,{\rm d}\gamma,
\end{align*}
where $f_\Gamma(\gamma; t) = -\frac{\partial}{\partial\gamma}P_{\rm C}(\gamma, t)$ is the probability density function (pdf) of ${\rm SINR}(t)$. Writing the $\log$ function in the integral form, we have
\begin{align*}
R(t) &= \int_{0}^\infty\int_{0}^\gamma \frac{1}{1+\omega}f_\Gamma(\gamma; t)\,{\rm d}\omega\,{\rm d}\gamma\\
&\hspace{-0.5cm}\overset{(a)}{=} \int_{0}^\infty \frac{1}{1+\omega}\left(\int_{\omega}^\infty f_\Gamma(\gamma; t)\,{\rm d}\gamma\right)\,{\rm d}\omega = \int_0^\infty \frac{P_{\rm C}(\gamma, t)}{1+\gamma}\,{\rm d}\gamma,
\end{align*}
where in $(a)$ we have changed the order of integration. This completes the proof.
\end{IEEEproof}
We now present the coverage probability and rate for the first service model. Based on Lemma \ref{lem:NoExclusion}, the network will remain homogeneous with density $\lambda_0$, and hence the coverage probability and rate of the network will follow that of a homogeneous PPP. Consequently, the coverage probability and rate of the first service model will be \eqref{MainCoverage} and \eqref{MainRates} evaluated at $t=0$, respectively, i.e.,
\begin{align}
P_{\rm C}(\gamma) &= \int_0^\infty 2\pi\lambda_0 u_0 {\rm e}^{-\pi \lambda_0 u_0^2} \times {\rm e}^{-\frac{\gamma \left(u_0^2+h^2\right)^{\alpha/2}}{P}N_0} \times \nonumber\\
&\exp\Big(- 2\pi\lambda_0{\displaystyle \int_{u_0}^\infty}\frac{u_\nbx}{1 + \frac{1}{\gamma}\left(\frac{u_\nbx^2+h^2}{u_0^2+h^2}\right)^{\alpha/2}}\,{\rm d}u_\nbx\Big) \,{\rm d}u_0, \label{SubMainCoverage}\\
R &= \int_0^\infty \frac{P_{\rm C}(\gamma)}{1+\gamma}\,{\rm d}\gamma. \label{SubMainRates}
\end{align}

\begin{figure}[t!]
    \centering
    \includegraphics[width=0.85\columnwidth]{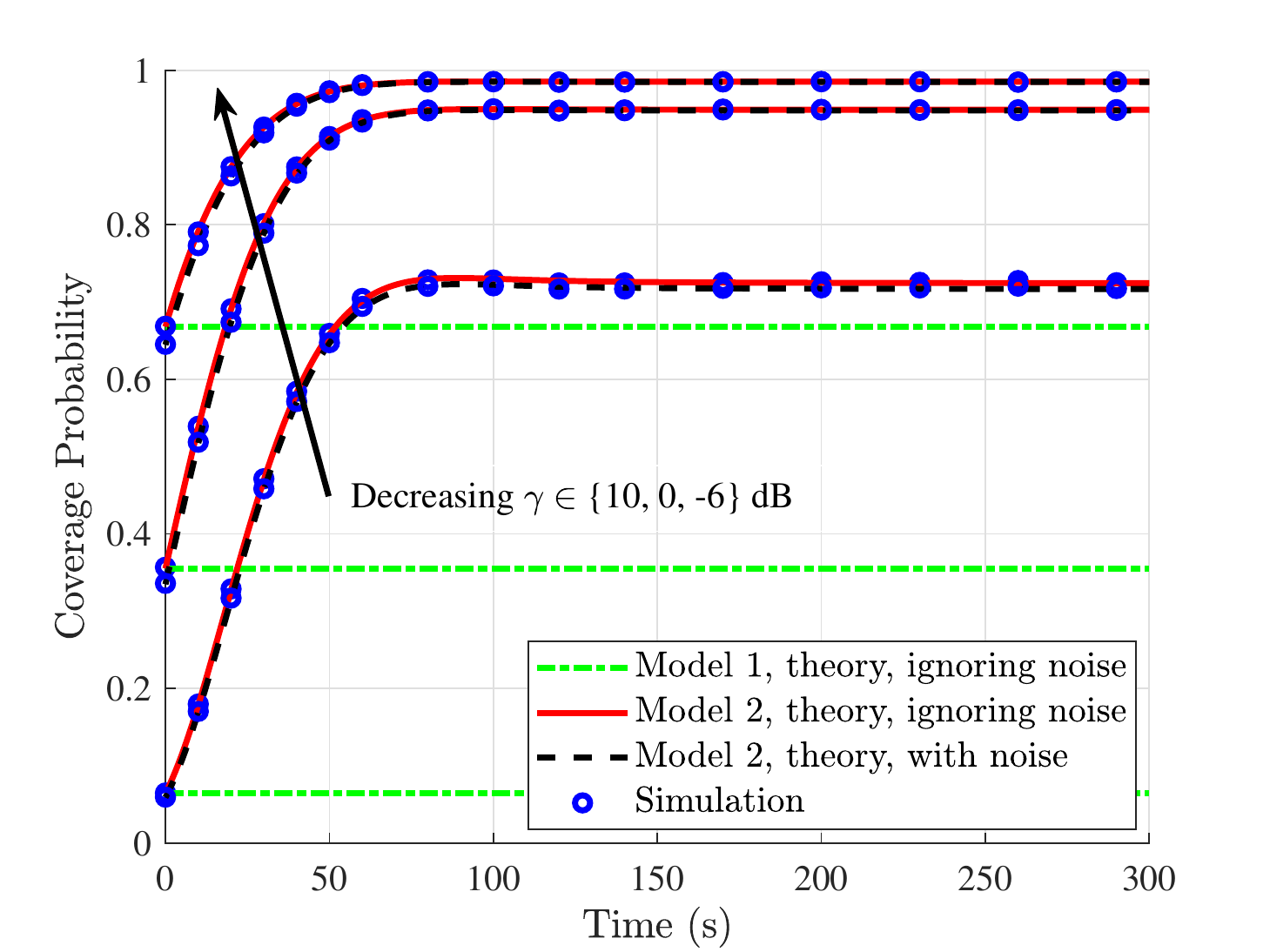}
    \caption{Time-varying coverage probability of the network for both service models. DBSs are moving at height $h = 100~{\rm m}$ and the path loss exponent is $\alpha = 3$.}\vspace{-0.3cm}
    \label{Fig:CoveragePlot}
\end{figure}

%%%%%%%%%%%%%%%%%%%%%%
\section{Numerical Results} \label{sec:Simulation}
%%%%%%%%%%%%%%%%%%%%%%
In this section, we numerically evaluate the theoretical expressions for the density of the network of interfering DBSs, the time-varying coverage probability, and the instantaneous received rate at the typical UE. Moreover, we analyze the impact of different network parameters on the density and rate. We assume that the initial locations of DBSs are distributed based on a PPP with density $\lambda_0 = 10^{-6}$ at height $h$ from the ground on $b(\nbo', R_{\rm D})$, where $R_{\rm D} = 100~{\rm km}$. For low altitude platform (LAP) DBSs, we assume $h \in \{100, \, 200\}$ meters as two typical values for the height, in which DBSs move at a constant velocity of $v = 45~{\rm km/h}$ at this height. Path loss exponent is assumed to take values $\alpha \in \{2.5, 3, 3.5\}$ depending on the environment. To demonstrate the effect of noise in our model, we performed the simulations both with and without noise. In order to set the noise power, we need to dimension the system appropriately. One way of doing that is to first introduce the concept of a cell-edge user, which is defined as the one located at distance $d_{\rm Edge}$ from the serving DBS, where $d_{\rm Edge}$ is the limiting value of $d$ in the equation $\nbbP[r_0 > d] \leq P_{\rm Edge}$ and we set $P_{\rm Edge} = 0.05$ in this illustration. Now, we can set the noise power $N_0$ such that the received signal-to-noise ratio (${\rm SNR}$) at a cell-edge user is some predefined value, which we set to $0{\rm ~ dB}$. Hence, assuming that the DBS transmit power is $P=0{\rm ~ dB}$, we set the noise power to be $N_0 = \left( h^2 - \frac{\log(P_{\rm Edge})}{\pi\lambda_0} \right)^{-\alpha/2}$.

Fig. \ref{Fig:DensityPlot} shows the evolution of the network density over time for the second service model, with an exclusion zone radius of $u_0 = 500~{\rm m}$ and four time instances $t \in \{20, 40, 50, 200\}$ seconds. As emphasized in the previous section, one interesting fact about the density of the interference field is that after time $t = \frac{u_0}{v}$, the density will have two homogeneous parts ($\lambda=\lambda_0$) and one bowl-shaped inhomogeneous part. As $t\to\infty$, the inhomogeneous part shrinks to $\lambda_0$, which eventually makes the interference field homogeneous.

In Fig. \ref{Fig:CoveragePlot}, we plot the coverage probability of the network for both service models at various values of ${\rm SINR}$ threshold as a function of time, both with and without including noise. As it is clear in this figure, coverage probability of the second service model will saturate quite fast after the DBSs start to move. Furthermore, the plots for the first service model are provided here to highlight the advantage of the second service model over the first one. The comparison of the results with and without noise demonstrates that the considered setup is interference limited.

\begin{figure}[t!]
    \centering
    \includegraphics[width=0.85\columnwidth]{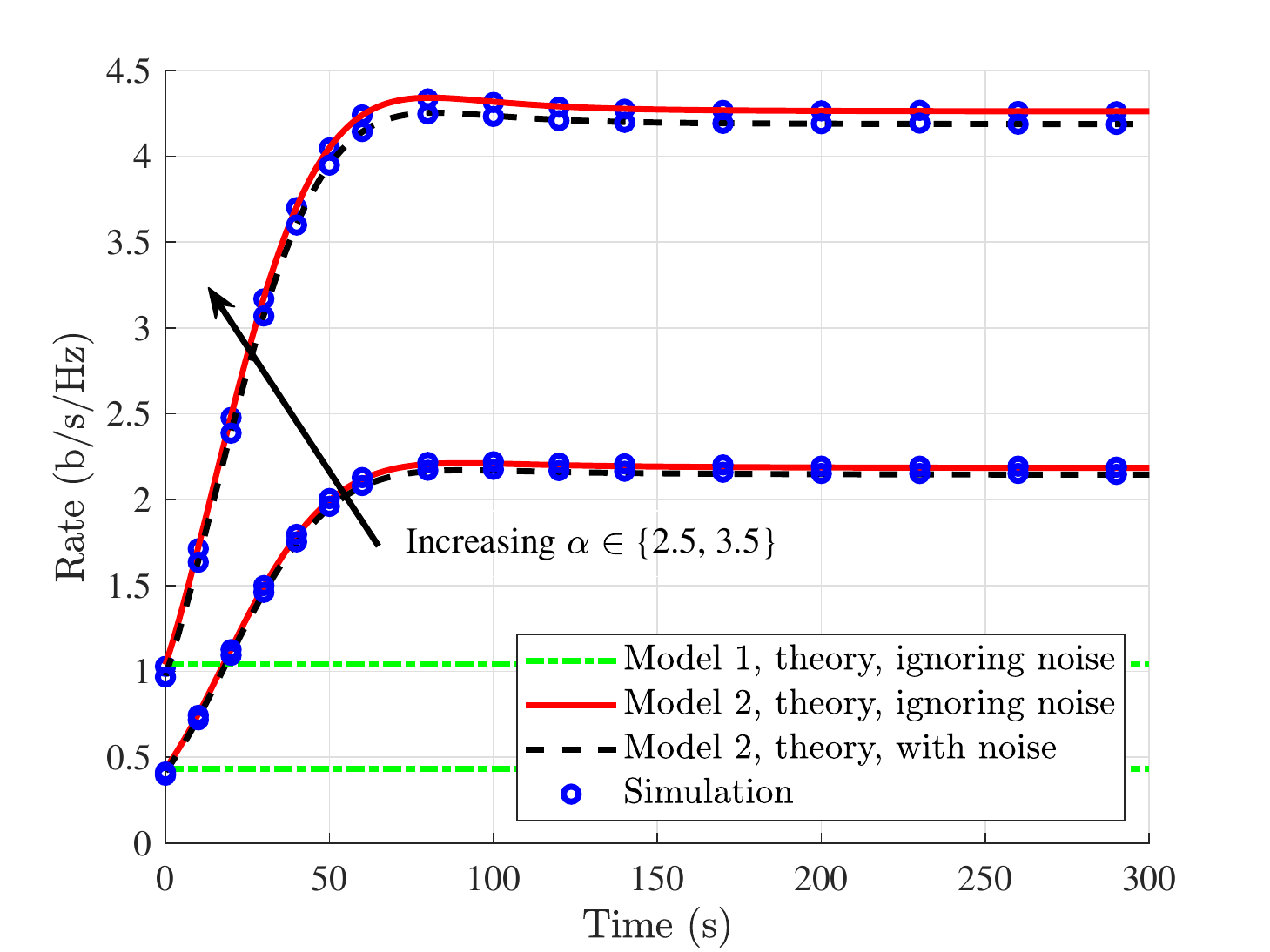}
    \caption{Time-varying rate of the network for both service models for different values of the path loss exponent, i.e., $\alpha \in \{2.5, 3.5\}$ and when DBSs are moving at height $h = 100~{\rm m}$.}\vspace{-0.3cm}
    \label{Fig:RatePlot_VaryAlpha}
\end{figure}

\begin{figure}[t!]
    \centering
    \includegraphics[width=0.85\columnwidth]{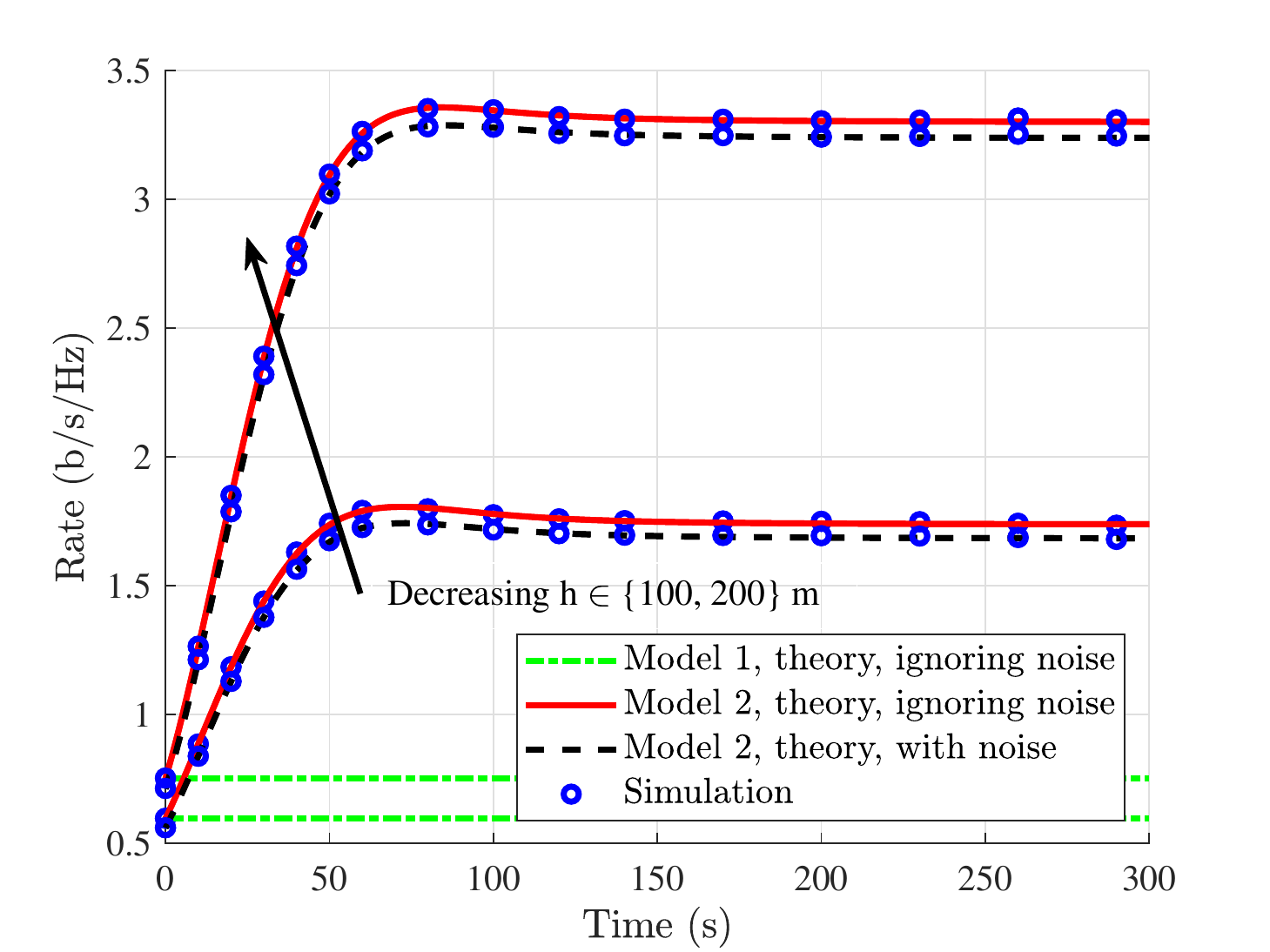}
    \caption{Time-varying rate of the network for both service models when DBSs are moving at height $h \in \{100, 200\}$ meters and the path loss exponent is $\alpha = 3$.}\vspace{-0.3cm}
    \label{Fig:RatePlot_VaryHeight}
\end{figure}

Analytic and simulation results for the time-varying rate of the network for both service models are provided in Figures \ref{Fig:RatePlot_VaryAlpha} and \ref{Fig:RatePlot_VaryHeight}. We compare the received rate at the typical UE when DBSs are at height $h=100~{\rm m}$ with various path loss exponents to see the effect of channel in Fig. \ref{Fig:RatePlot_VaryAlpha}. As it is clear in this figure, as $\alpha$ increases, the received rate will also increase. This is mainly due to the fact that as $\alpha$ increases, although the received power at the typical UE will decrease, the interference power decreases as well, which cumulatively increases the received rate. As in Fig. \ref{Fig:CoveragePlot}, we again compare the results for our second service model with the first model and draw the same general conclusions.

To observe the effect of height in the received rate, we plot the received rate at various heights with the same path loss exponent in Fig. \ref{Fig:RatePlot_VaryHeight}. We can see in this figure that the received rate will decrease as height increases. This interesting fact could have been observed directly from \eqref{MainCoverage} and \eqref{MainRates} as well.

\vspace{-0.2cm}
%%%%%%%%%%%%%%%%%%%%%%
\section{Conclusion} \label{sec:Conclusion}
%%%%%%%%%%%%%%%%%%%%%%
In this paper, we modeled the mobility of a network of DBSs, which are initially distributed as a PPP at a constant height from the ground. The serving DBS is selected based on a nearest neighbor association policy and moves according to two specific service models, i.e., (i) in a random direction, and (ii) towards the typical UE and hovers above the location of the UE. Inspired by the simulation models used by 3GPP, we assumed that all the other DBSs (treated as interfering DBSs) move on straight lines and in random directions, independent of the serving DBS. In both service models, the network of interfering DBSs will be inhomogeneous PPPs whose densities are derived using displacement theorem. We then compared our mobility model with other mobility models in which DBSs can also change their direction of movement and demonstrated that our straight-line model can be regarded as a lower bound on the system performance to these mobility models. We finally analyzed the time-varying coverage probability and rate of the network for both service models. Mathematical analysis of more complex mobility models, such as RW and RWP are left as future work. To the best of our knowledge, this is the first work that analyzes a 3GPP-inspired drone mobility model and connects it to more general non-linear mobility models.

\bibliographystyle{IEEEtran}
\bibliography{C1_3GPP_SG_Mobility_Drone}

\end{document}